\documentstyle[12pt,epsf,psfig]{article}
\def\J{$J/\psi$}
\def\j{J/\psi}
\def\X{$\chi$}

\def\P{$\psi'$}

\def\be{\begin{equation}}
\def\ee{\end{equation}}

\def\lsim{\raise0.3ex\hbox{$<$\kern-0.75em\raise-1.1ex\hbox{$\sim$}}}
\def\gsim{\raise0.3ex\hbox{$>$\kern-0.75em\raise-1.1ex\hbox{$\sim$}}}

\def\NP{{ Nucl.\ Phys.\ }}
\def\PL{{ Phys.\ Lett.\ }}
\def\PR{{ Phys.\ Rev.\ }}

\def\PRL{{ Phys.\ Rev.\ Lett.\ }}

\def\ZP{{ Z.\ Phys.\ }}

\begin{document}

\noindent June 8, 1998 \hfill BI-TP 98/16

\vskip 1.0cm

\centerline
{\Large \bf A Brief History of \J~Suppression\footnote{Invited talk
at the RHIC/INT Workshop {\sl `Quarkonium Production in Relativistic
Nuclear Collisions'}, Institute for Nuclear Theory, Seattle, Washington,
USA, May 11 - 15, 1998.}}

\vskip 0.8cm

\centerline{\bf Helmut Satz}

\bigskip

\centerline{Fakult\"at f\"ur Physik, Universit\"at Bielefeld}

\par

\centerline{D-33501 Bielefeld, Germany}

\vskip 1.0cm

Statistical QCD predicts that strongly interacting matter will become
deconfined at high temperatures and/or densities. The aim of high
energy nuclear collisions is to study the onset of deconfinement and
the properties of deconfined media in the laboratory. Hence it is
essential to define an unambiguous and experimentally viable probe for
deconfinement. Twelve years ago, T.\ Matsui and I proposed that
\J~production should constitute such a probe \cite{M&S}, and I want to 
sketch here rather briefly the evolution of this idea in the light of
subsequent experimental and theoretical work. To keep my report really
brief, I will concentrate on what seems to me the main line of
development, with sincere apologies to all those whose work is not
adequately considered or mentioned.

\medskip

Matsui and I had argued that in a quark-gluon plasma colour screening
dissolves the \J~into its $c$ and $\bar c$ constituents, which
separate and thus at hadronisation have to combine
with light quarks to form a $D$ and a $\bar D$ instead of a \J.
Since the overall Drell-Yan dilepton rates remain unaffected,
deconfinement must lead to a suppression of \J~production
relative to that of Drell-Yan pairs. Hotter media dissolve more
\J's, so the suppression should increase with centrality.

\medskip

Less than a year later, the NA38 collaboration at the CERN SPS
observed in $O-U$ interactions a considerable
suppression of \J~production, increasing with the centrality of the
collision \cite{NA38Mo,NA38OU}. This result was subsequently corroborated in
$S-U$ collisions \cite{NA38SU}.

\medskip

Since \J~production was known to be attenuated also in $p-A$ collisions
\cite{NA3}, it was natural to ask if the observed $O-U$ and $S-U$
suppression could be understood simply in terms of a \J~absorption
in the nuclear matter of target and projectile \cite{Capella,G-H-1}.
With a value of $\sigma^{\rm in}_{\j-N} \simeq 4$ mb as determined by the
$p-A$ data of \cite{NA3}, nuclear absorption alone was found to to produce
considerably less \J~suppression than observed in $O-U$ and $S-U$ collisions.
On the other hand, nucleus-nucleus collisions lead to
the abundant production of hadronic secondaries, and such `comovers'
could also dissociate \J's \cite{B-M} - \cite{Ftac2}. Subsequently, a
suitable combination of absorption on nucleons and comovers was
indeed shown to reproduce the observed suppression \cite{Gavin,GVTS},
indicating that dense hadronic matter alone, without any deconfinement,
would be sufficient.

\medskip

It thus became necessary to look for features which distinguish between
absorption and deconfinement as suppression mechanisms. The main difference
is obviously the sudden onset of deconfinement as a critical
phenomenon, in contrast to absorption as an effect always present and
hence increasing gradually with density. The well-defined onset of
deconfinement moreover leads to a characteristic step-wise suppression
pattern for the measured \J's. It is known that only about 60 \% of
these are produced directly as 1S states; of the remainder, about
30 \% come from \X~and about 10 \% from \P~decay \cite{Lemoigne,Anton}. Colour
screening studies of charmonium dissociation show that the larger
excited states are dissolved at lower densities than the tightly
bound ground state \J~\cite{MTM,K-S}. In particular, the \X~and the
\P~are expected to `melt' at the deconfinement point, while the direct
\J~needs an energy density about twice as high. As a result,
deconfinement leads for the measured \J's to the successive suppression
pattern shown in Fig.\ 1, in contrast to the monotonic decrease of
the survival probability obtained from absorption. The generic step
pattern shown here will of course be softened by nuclear profile
effects, impact parameter uncertainties, etc.; on the other hand, this
could be partially compensated if there is a really discontinuous onset of
deconfinement as function of the energy density of the medium.

\begin{figure}[h]
\vspace*{-0mm}
\centerline{\psfig{file=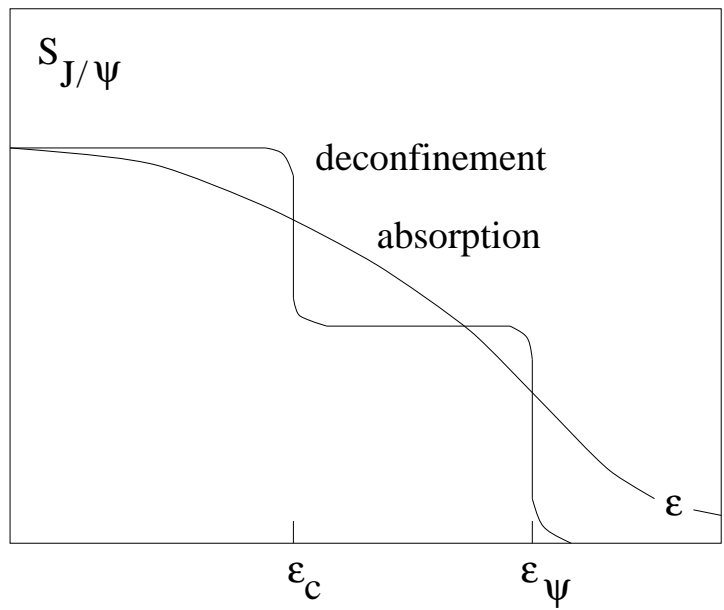,height= 80mm,angle= -90}}
\caption{The \J~survival probability $S_{\j}$ as function of the energy
density $\epsilon$, for suppression by deconfinement and by hadronic 
comover absorption.}
\end{figure}

As a sideline, we note that the observed \J~attenuation occurs mainly
at low transverse momentum \cite{NA38pt}. It was first thought that high
$p_T$ \J's could `escape' in space or time from suppression by
deconfinement \cite{K-P}, so that the $p_T$-dependence could be used to
determine size or life-time of the suppressing medium \cite{B-Opt}.
However, subsequent studies showed that initial state parton
interactions, prior to the formation of the $c \bar c$ pair, can in
fact account for all of the measured $p_T$ dependence of \J~suppression
\cite{GG} - \cite{B-Opt2}.

\medskip

The next step in the development came from Fermilab $p-A$ data, based on
four different nuclear targets \cite{E772}; these data confirmed the
nuclear attenuation previously observed \cite{NA3}, but in addition
they showed
the same amount of suppression for \J~and \P. This excluded the
interpretation of \J~suppression in $p-A$ interactions as absorption
of physical charmonium states, since the \P~is more than four
times larger than the \J, which should show up in the dissociation
cross sections \cite{Povh} and hence in nuclear absorption.
Nevertheless, using an `effective' dissociation cross section
of some 6 - 7 mb was found to account for all \J~suppression data
available at that time, from $p-A$ to central $S-U$ collisions
\cite{G-H-2}. Both the size of the required cross section and the
equality of \J~and \P~suppression remained as puzzles.

\medskip

The solution of these puzzles came from a combination of several
different theoretical and experimental developments. First, it was
rediscovered \cite{Kaidalov,KS3} that short distance QCD allows the
calculation of the cross section for the dissociation of heavy
quarkonia by incident light hadrons \cite{Peskin}. The calculation
is based on the known \J~parameters (binding energy, charm quark mass)
and on the gluon distribution function in hadrons as determined from
deep inelastic scattering. The resulting value of the dissociation cross
section for light hadrons of momenta $p_h \leq 3$ GeV/c incident on a
stationary \J~is found to lie by some $10^{-1}$ to $10^{-3}$ below the
geometric cross section of about 3 mb. Hence hadrons in this momentum
range cannot really break up a \J. On the other hand, dissociation by
gluons (the QCD version of the photo-effect) peaks for gluon momenta
around 0.8 - 1.0 GeV/c, making a quark-gluon plasma of temperatures $T
\geq 250$ MeV very effective for \J~dissociation. If we are provided
with physical charmonium states and a strongly interacting medium, the 
fate of the \J~thus becomes an unambiguous deconfinement test \cite{KS3}.

\medskip

The calculated extremely small dissociation cross section for slow
light hadrons incident on \J's effectively rules out suppression by
hadronic comovers at present energies. However, the short distance QCD
calculations on which this conclusion is based become exact only in
the limit of large heavy quark mass. It must therefore be checked
experimentally if the charm quark is already heavy enough to apply such
considerations, and such a check is indeed possible \cite{KS5}.

\medskip

Besides the threshold behaviour, short distance QCD also provides the
high energy value of the \J~dissociation cross section; the result of some
2 - 3 mb indicates once more that something else must be absorbed in the
mentioned $p-A$, $O-U$ and $S-U$ data.

\medskip

On the experimental side, a crucial contribution came from a study of
high $p_T$ charmonium production in $p \bar p$ interactions at the
Fermilab Tevatron \cite{CDF}. In such collisions, charmonium
production was believed to be accountable in terms of the colour
singlet model \cite{csm}: `perturbative' parton interactions produce a
$c \bar c$ pair in a colour octet state, which subsequently neutralizes
its colour and forms a colour singlet $c \bar c$ of \J~quantum numbers
by the emission of hard gluons. The model definitely disagreed with low
$p_T$ data on charmonium production \cite{Anton,Gavai}, but in this
kinematic regime it was not really expected to be reliable. The
Fermilab data, in contrast, were in a suitable kinematic region and
nevertheless turned out to disagree with the colour singlet model
predictions by factors 10 - 100. This clearly ruled out such a
description and showed that non-perturbative effects are important in
charmonium production.

\medskip

Enter: the colour octet model. Based on the formalism of
non-relativistic QCD \cite{BBL}, it assumes that in order to form a
charmonium state, the perturbatively produced $c\bar c$ pair first
neutralizes its colour by combining with one or more soft collinear
gluon comovers \cite{Braaten}. This pre-resonance charmonium state, a
colour singlet formed by the $c \bar c$ colour octet plus the smallest
number of gluons required to provide the
relevant quantum numbers, then is quickly transformed into the basic
colour singlet $c \bar c$ charmonium resonance. For the hadroproduction
of \J's or \P's at small transverse momentum, the lowest possible
pre-resonance states of this type are colour singlets of the octet $c
\bar c$ and one gluon; such a state turns into the basic colour singlet
$c \bar c$ states in a time of about 0.2 - 0.3 fm \cite{KS6}. The
charmonia produced in present $p-A$ data, i.\ e., around $x_F \simeq
0.1$, traverse the entire nucleus during the pre-resonance time, so that
the nuclear medium `sees' only the passage of the ${c \bar c}-g$ state.
Since this is of the same structure for \J~and \P, these states also
suffer the same absorption during their passage. Moreover, the break-up
of a state consisting of colour octet components leads to a cross 
section which is 9/4 times larger than that of a compact state
consisting of colour triplet components; this suggests some 5 - 7 mb as
the absorption cross section for the $c{\bar c}-g$ in nuclear matter, in
accord with the result of the phenomenological fit in \cite{G-H-2}.

\medskip

Pre-resonance charmonium absorption thus provides the needed explanation
for the required size of the dissociation cross section as well as for
the equal absorption suffered by the different charmonium states
\cite{KS3}. It also removes the necessity for any additional absorption
by hadronic comovers in $O-U$ or $S-U$ collisions; pre-resonance
absorption in nuclear matter alone reproduces well all \J~suppression
from $p-A$ to central $S-U$ data, including the centrality dependence
of nucleus-nucleus collisions \cite{KLNS}. The absence of comover
suppression for the \J, however, does not imply the absence of hadronic
comovers. This is shown by the fact that the loosely bound and hence
easily broken \P~does in fact suffer more than just pre-resonance
suppression in central $S-U$ interactions \cite{NA38psi'}. So hadronic
comovers appear, but they do not affect the \J, in accord with the
mentioned short distance QCD calculations \cite{KS3}. The conclusion
at the end of the `sulphur era' thus found the observed \J~suppression,
from $p-A$ to $S-U$ collisions, to be `normal': it could be understood
in terms of `conventional' pre-resonance absorption in nuclear matter
and required no additional suppression by hadronic comovers or by 
colour deconfinement \cite{KS6} - \cite{Ramona}.

\medskip

The data from the 1995 $Pb-Pb$ run of NA50 marked the end of this normality
and the advent of `anomalous' \J~suppression \cite{NA50Mo,NA50Hei}:
increasing with centrality, the \J~data fell about 30 \% below the
value expected from pre-resonance absorption, indicating the onset of a
new suppression mechanism. Such an additional decrease would in fact
occur if all \J~in the hot inner collision region would `melt'
\cite{Gupta,BO97}, with the energy density in central $S-U$ collisions 
taken as critical deconfinement
threshold. Explanations in terms of absorption by hadronic comovers were
also considered once more \cite{G-V,C-A}; since any such
absorption sets in gradually, this required some effect to be present already
in $S-U$ data, where it was not really needed \cite{KLNS}. Nevertheless,
by admitting less than optimal fits to $p-A$ and $S-U$ data, hadronic
comover absorption remained an alternative to be considered. In
particular, the rather crucial relation between central $S-U$ and
peripheral $Pb-Pb$ data still remained unclear.

\medskip

A year later, the much more precise data of the 1996 $Pb-Pb$ run
appeared \cite{Gonin} - \cite{NA50Mo'}, showing peripheral $Pb-Pb$ data in
accord with pre-resonance absorption (and thus with the $S-U$ results).
But in addition, these data indicated with increasing centrality an
abrupt onset of the anomalous suppression (Fig.\ 2): at an impact
parameter of about 8 fm (corresponding to an associated transverse
energy of about 40 GeV), the ratio of \J~production to that of
Drell-Yan pairs suddenly dropped by about 25 \%.

\begin{figure}[h]
\centerline{\psfig{file=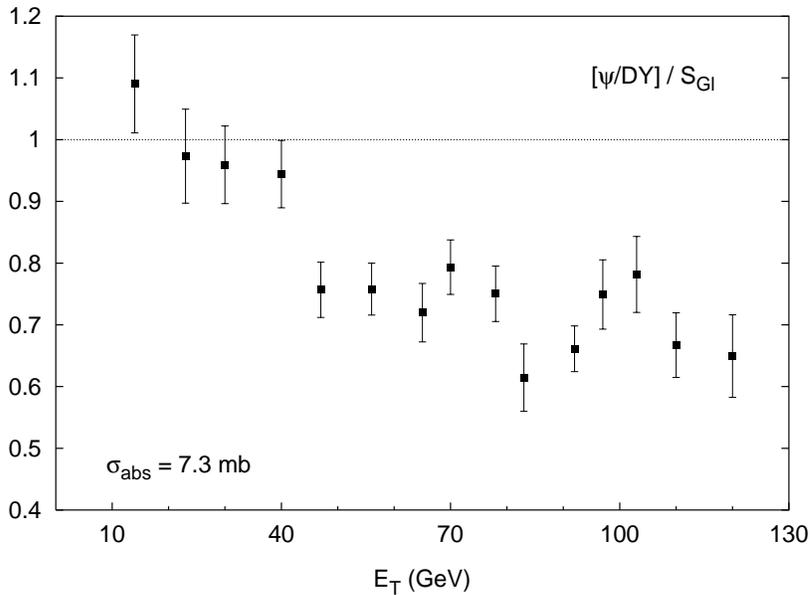,height= 80mm,angle=0}}
\caption{Anomalous \J~suppression relative to Drell-Yan production 
\cite{NA50Mo'}, normalized to pre-resonance nuclear absorption ($S_{\rm Gl}$)
with a cross section of 7.3 mb \cite{KLNS}.}
\end{figure}

\medskip

Since an abrupt change of physical variables is in general related to
critical behaviour, this observation might well be the first indication 
of deconfinement; hence it quite naturally triggered numerous
questions. Are the \J~and the Drell-Yan data in the different kinematic
regions absolutely reliable? If yes, how abrupt is the onset, and
how does it show up as function of other variables? How abrupt can it
realistically be, given impact parameter smearing, resolution, etc.?
How precise do data have to be to rule out models based on a smooth
transition? Which variable governs the onset of anomalous suppression?
What further data are best suited to corroborate and eventually confirm the
observation? -- to name just some of the many questions. Clearly,
the primary objective of the experimentalists must be to establish their
results beyond any reasonable doubt.

\medskip

Until that is done, perhaps the best a theorist can do is to go
back to basics and ask what QCD can tell us in more detail about the
phenomena expected to occur at the onset of deconfinement. Some first
work in this direction, studying the relation of deconfinement and
percolation, is reported elsewhere \cite{perco}; its application
to \J~suppression will be presented at this meeting by M.\ Nardi
\cite{nardi}.

\end{document}